%Paper: hep-th/9410226
%From: Simon Brian Davis <S.B.Davis@damtp.cambridge.ac.uk>
%Date: Fri, 28 Oct 94 23:45 GMT
%Date (revised): Sat, 29 Oct 94 00:11 GMT

\magnification=1200
\parskip 10pt plus 5pt
\parindent 14pt
\baselineskip=18pt
\input mssymb
\pageno=0
\footline={\ifnum \pageno <1 \else \hss \folio \hss \fi}
\line{\hfil{DAMTP-R/94/29}}
\line{\hfil{August, 1994}}
\vskip 1in
\centerline{\bf THE BOSONIC STRING MEASURE AT TWO AND THREE LOOPS AND}
\vskip 1pt
\centerline{\bf SYMPLECTIC TRANSFORMATIONS OF THE VOLUME FORM}
\vskip .65in
\centerline{Simon Davis}
\vskip .4in
\centerline{Department of Applied Mathematics and Theoretical Physics}
\vskip 1pt
\centerline{University of Cambridge}
\vskip 1pt
\centerline{Silver Street, Cambridge CB3 9EW}
\vskip .5in
\noindent{\bf Abstract.} Symplectic modular invariance of the bosonic string
partition function has been verified at genus 2 and 3 using the period matrix
coordinatization of
moduli space.  A calculation of the transformation of the holomorphic part
of the differential volume element shows that an extra phase
arises together with the factor associated with a specific modular weight;
the phase is cancelled in the transformation of the entire volume element
including the complex conjugate.  An argument is given for modular invariance
of the reggeon measure at genus twelve.
\vfill
\eject

The formulas for the bosonic string partition function at two and three loops
have been given as integrals of the type
$$Z_g~=~\int_{{\cal M}_g}~\prod_{i=1}^{dim~{\cal M}_g}~dy_i \wedge d{\bar y}_i
{}~\vert F(y_i)\vert^2~[ det(Im~\tau)]^{-13}
\eqno(1)$$
where the coordinates $y_i$ may be represented by elements of the period matrix
$\tau$ and
$F(y_i)$ represents a section of a holomorphic bundle over moduli space [1].
The restriction to a single copy of moduli space is equivalent to integration
over a fundamental domain, in the subspace of period matrices in the
${1\over 2} g(g+1)$-dimensional Siegel upper half space ${\cal H}_g$
[$det(Im~\tau) >0$]
corresponding to Riemann surfaces, defined with respect to the symplectic
modular group Sp(2g; ${\Bbb Z}$)
$$\tau~\to~ (A\tau+B)(C \tau+D)^{-1}~~~~~~\left(\matrix{A&B \cr
                                                         C&D\cr}\right)
\in Sp(2g;{\Bbb Z})
\eqno(2)$$
Then the function $F(y_i)$ can be expressed as a modular function of the
period matrix [2][3][4], implying
$$\eqalign{Z_2~&=~\int_{{\cal M}_2}~\prod_{i=1}^2~d\tau_{ij} \wedge d {\bar
\tau}_{ij}~
\vert \chi_{10} (\tau)\vert^{-2}~ [det(Im~\tau)]^{-13}
\cr
Z_3~&=~\int_{{\cal M}_3}\prod_{i=1}^3~d \tau_{ij}\wedge d{\bar \tau}_{ij}~\vert
\chi_{18}(\tau)
\vert^{-1}~[det(Im~\tau)]^{-13}
\cr}
\eqno(3)$$
Given a symplectic transformation of the period matrix
$$\eqalign{Im~\tau~&\to~{1\over {2i}}[(A \tau+B)(C\tau+D)^{-1}-(A{\bar \tau}+B)
(C{\bar \tau}+D)^{-1}]
\cr
&~~~~=~{1\over {2i}} (\tau C^T+D^T)^{-1}[(\tau A^T+B^T)(C {\bar \tau}+D)
{}~-~(\tau C^T +D^T)(A {\bar \tau}+B)]
\cr
&~~~~~~~~~~~~~~~~~~~~~~~~~~~~~~~~~~~~~~~~~~~~~~~~~~~~~~~~~
{}~~~~~~~~~~~~~~~~~~~~~~~~~~~~\cdot (C{\bar \tau}+D)^{-1}
\cr
&~~~~=~(\tau C^T + D^T)^{-1}~(Im~\tau)~(C {\bar \tau} +D)^{-1}
\cr}
\eqno(4)$$
so that
$$ [det (Im~\tau)]^{-13}~\to~[det (Im~\tau)]^{-13} \cdot \vert det(C \tau
+D)\vert^{26}
\eqno(5)$$
Since
$$\vert \chi_{10} (\tau)\vert^{-2}~\to~\vert \chi_{10}(\tau)\vert^{-2}
{}~\vert det(C \tau + D)\vert^{-20}
\eqno(6)$$
modular invariance of the integrand in $Z_2$ depends on the transformation
properties of the differential volume element.
At genus two, the Jacobian of the coordinate transformation from $(\tau_{11},
\tau_{12}, \tau_{22})$ to
$(\tau_{11}^\prime,\tau_{12}^\prime,\tau_{22}^\prime)$,
 $det\left({{\partial (\tau_{11}^\prime,\tau_{12}^\prime,
\tau_{22}^\prime)}\over
{\partial(\tau_{11},\tau_{12}, \tau_{22})}}\right)$, where
\hfil\break
${{\partial
 \tau_{ij}^\prime}\over {\partial
\tau_{kk}}}~=~[A-(A\tau+B)(C\tau+D)^{-1}C]_{ik} (C\tau+D)_{kj}^{-1}$
and  ${{\partial \tau_{ij}^\prime}\over {\partial \tau_{12}}}~=~
[A-(A\tau+B)(C\tau+D)^{-1}]_{i1}(C \tau+D)_{2j}^{-1}
+[A - (A\tau+B)(C\tau+D)^{-1}C]_{i2} (C\tau+D)_{1j}^{-1}$,
equals
$$\eqalign{det &\left(\matrix{ [A-(A\tau+B)(C\tau+D)^{-1}
C]_{11}&[A-(A\tau+B)(C\tau+D)^{-1}C]_{12}\cr
(C\tau+D)_{12}^{-1}& (C\tau+D)_{22}^{-1}\cr}\right)
\cr
&~~~~~~~~~~det[A-(A\tau+B)(C\tau+D)^{-1}C]~det(C\tau+D)^{-1}
\cr
&~~~=~det\left(\matrix{[A-(A\tau+B)(C\tau+D)^{-1}C]_{11}&[A-(A\tau+B)
(C\tau+D)^{-1}C]_{12}\cr
(C\tau+D)_{12}^{-1}& (C\tau+D)_{22}^{-1}\cr}\right)
\cr
&~~~~~~~~~~~~~det(C\tau+D)^{-2}
\cr}
\eqno(7)$$
as $det(A-(A\tau+B)(C\tau+D)^{-1}C)=det(AC^{-1}D-B)~det(C\tau+D)^{-1}~det C
=det(C\tau+D)^{-1}$.  Moreover, equality of the differentials
$d\tau_{12}^\prime$ and $d\tau_{21}^\prime$ implies that
$$\eqalign{{{[A-(A\tau+B)(C\tau+D)^{-1}C]_{11}}\over {(C\tau+D)_{11}^{-1}}}
{}~&=~{{[A-(A\tau+B)(C\tau+D)^{-1}C]_{12}}\over {(C\tau+D)_{21}^{-1}}}
\cr
{}~&=~{{[A-(A\tau+B)(C\tau+D)^{-1}C]_{21}}\over {(C\tau+D)_{12}^{-1}}}
\cr
{}~&=~{{[A-(A\tau+B)(C\tau+D)^{-1}C]_{22}}\over {(C\tau+D)_{22}^{-1}}}
\cr}
\eqno(8)$$
and the Jacobian becomes
$${{[A-(A\tau+B)(C\tau+D)^{-1}C]_{11}}\over {(C\tau+D)_{11}^{-1}}}
{}~det(C\tau+D)^{-3}
\equiv \Phi_2(\tau)~ det(C\tau+D)^{-3}
\eqno(9)$$
The transformation of the holomorphic part of the differential volume
element is then $\prod_{i\le j =1}^2~d\tau_{ij}~\to~\prod_{i\le j=1}^2
d\tau_{ij}
\Phi_2(\tau) det(C\tau+D)^{-3}$.  The factor $\Phi_2(\tau)$ represents a
modification of earlier calculations of the transformations of the holomorphic
part of the volume element [3].
It can be shown, however, that $\Phi_2(\tau)$ is a phase factor, because
$$A-(A\tau+B)(C\tau+D)^{-1}C~=~\Phi_2(\tau) [(C\tau+D)^{-1}]^T
\eqno(10)$$
from the relations (8) and the $\Phi_2(\tau)^2=1$ from the equality of the
determinants.  Consequently, the phase factor is cancelled by the complex
conjugate in the transformation of $\prod_{i\le j =1}^2~d {\bar \tau}_{ij}$ and
$$\prod_{i\le j =1}^2~d\tau_{ij} \wedge d{\bar\tau}_{ij}
{}~\to~\prod_{i\le j=1}^2~d\tau_{ij} \wedge d {\bar \tau}_{ij}~\vert
det(C\tau+D)
\vert^{-6}
\eqno(11)$$
Equations (5), (6) and (11) give the required modular invariance of the
two-loop bosonic string partition function.

At three loops, the transformation from the coordinates  $\{(\tau_{ij}),~1\le
i\le j\le 3\}$ to $\{(\tau_{ij}^\prime),~1 \le i\le j \le 3\}$ gives rise to
a Jacobian involving the sum of terms of the form
$$\prod_{\#[ij]=6}~[A-(A\tau+B)(C\tau+D)^{-1}C]_{ij}~
\prod_{\#[kl]=6}~(C\tau+D)_{kl}^{-1}
\eqno(12)$$
Since $\tau$ is a $3\times 3$ matrix, one obtains
$$\prod_{i\le j=1}^3~d\tau_{ij}~\to~\prod_{i\le j=1}^3~d\tau_{ij}~
\Phi_3(\tau)~det(C\tau+D)^{-4}
\eqno(13)$$
for the transformation of the holomorphic part of the differential volume
element, where $\Phi_3(\tau)$ is a phase factor.  Again, the phase factor
represents a modification of the transformation rule which is cancelled by
its complex conjugate in the action of the symplectic modular group on the
total differential volume element
$$\prod_{i\le j=1}^3~d\tau_{ij} \wedge d{\bar\tau}_{ij}~\to~
\prod_{i\le j=1}^3~d\tau_{ij} \wedge d{\bar\tau}_{ij}~\vert det(C\tau+D)
\vert^{-8}
\eqno(14)$$
Since
$$\vert\chi_{18}(\tau)\vert^{-1}~\to~\vert\chi_{18}(\tau)\vert^{-1}
{}~\vert det(C\tau+D)\vert^{-18}
\eqno(15)$$
modular invariance of the string integrand at three loops follows from
equations
(5), (14) and (15).

Thus, although a phase factor arises in the transformation of the holomorphic
part of the differential volume element, it does not affect modular invariance
of the measure.  Moreover, equations (11) and (14) are special cases of a
formula
which holds for arbitrary genus.
$$\prod_{i\le j=1}^g~d\tau_{ij} \wedge d{\bar \tau}_{ij}
{}~\to~{{\prod_{i\le j=1}^g~ d\tau_{ij}\wedge d{\bar \tau}_{ij}}\over
{\vert det(C\tau+D)\vert^{2(g+1)}}}
\eqno(16)$$
follows from the invariance of the volume form
$${{\bigwedge_{(i,j)=1}^{{1\over 2}g(g+1)}~d\tau_{ij} \wedge d{\bar\tau}_{ij}}
\over {[det(Im~\tau)_{ij}]^{g+1}}}
\eqno(17)$$
which may be deduced from the symplectic hermitian metric [5]
$$ds^2~=~trace[(Im~\tau_{ij})^{-1} d\tau_{ij} (Im~\tau_{ij})^{-1}
d{\bar \tau}_{ij}]
\eqno(18)$$
on the Siegel upper half space ${\cal H}_g$.

The volume form (17) can be obtained at low genus by pulling back the
Hodge norm on $E^{g+1}$, where $E$ is the Hodge line bundle over ${\cal M}_g$,
to $K$, the determinant line bundle of the holomorphic cotangent bundle,
using the isomorphism $K\simeq E^{g+1}$ valid for ${\cal M}_1,~{\cal M}_2$, and
${\cal M}_3^0={\cal M}_3-{\it h}_3$, with ${\it h}_3$ being the subvariety of
genus-three hyperelliptic surfaces [5].

When $g>3$, the dimension of the Siegel upper half space is greater than
that of the moduli space ${\cal M}_g^0$, but a modular invariant measure on
${\cal M}_g^0$
can be defined by using the volume form resulting from the metric induced
by the embedding  of ${\cal M}_g^0$ into ${\cal H}_g$.  It has already been
noted that as
the relation
$$\tau_{mn}~=~{1\over {2\pi i}}~\left[~ln~K_n~\delta_{mn}
{}~+~\sum_\alpha~^{(m,n)}~ln~\left({{\xi_{1m}-V_\alpha\xi_{1n}}\over
{\xi_{1m}-V_\alpha\xi_{2n}}}{{\xi_{2m}-V_\alpha \xi_{2n}}\over
{\xi_{2m}-V_\alpha \xi_{1n}}}\right)~\right]
\eqno(19)$$
holds for any surface which can be uniformized by a Schottky group, it provides
a coordinatization of the Schottky locus using the variables
$K_n,~\xi_{1n}$ and $\xi_{2n}$ [6]. In particular, period matrices of the form
(19),
with restrictions placed on multipliers and fixed points to avoid
overlapping of isometric circles,  should provide
solutions of the KP equation
$${3\over 4}u_{yy}~=~{\partial \over {\partial x}} \left(u_t-{1\over 4}u_{xxx}
-{3\over 2}uu_x \right)
\eqno(20)$$
through
$$u(x,y,t)~=~2{\partial_x}^2~log~\Theta(Ux+Vy+Wt+z_0~\vert~\tau)
\eqno(21)$$
for some three-dimensional vectors $U,~V,~W \in {\Bbb C}^g$ and any $z_0\in
{\Bbb C}^g$ [7][8][9][10].  The Schottky locus can be described also by
an infinite set of inequalities involving the period matrix defining the
intersection of the fundamental region of the modular group $Sp(2g;{\Bbb Z})$
with the space of Riemann surfaces.  At large genus, it has been shown that
this set of inequalities can be reduced to a finite set in the degeneration
limit $\vert K_n\vert \to 0$ leading to an exhaustion of the fundamental
regions as $g$ increases [11].

Using (19), the metric (18) can be re-expressed in terms of the Schottky
group parameters.
Since
$$\eqalign{d\tau_{nn}~=~{1\over {2\pi i}}~\biggl[~{{dK_n}\over {K_n}}~&+~
\sum_\alpha~^{(n,n)}\sum_m~{d\over {dK_m}}~ln~\left({{\xi_{1n}-V_\alpha
\xi_{1n}}
\over {\xi_{1n}-V_\alpha \xi_{2n}}}{{\xi_{2n}-V_\alpha \xi_{2n}}\over
{\xi_{2n}-V_\alpha \xi_{1n}}}\right)~ dK_m
\cr
{}~&+~\sum_\alpha~^{(n,n)}\sum_m {d\over {d\xi_{1m}}}~ln~
\left({{\xi_{1n}-V_\alpha
\xi_{1n}}\over {\xi_{1n}-V_\alpha \xi_{2n}}} {{\xi_{2n}-V_\alpha \xi_{2n}}
\over {\xi_{2n}-V_\alpha \xi_{1n}}}\right)~ d\xi_{1m}
\cr
{}~&+~\sum_\alpha~^{(n,n)}\sum_m {d\over
{d\xi_{2m}}}~ln~\left({{\xi_{1n}-V_\alpha
\xi_{1n}}\over {\xi_{1n}-V_\alpha\xi_{2n}}}{{\xi_{2n}-V_\alpha \xi_{2n}}\over
{\xi_{2n}-V_\alpha \xi_{1n}}}\right)~ d\xi_{2m}~\biggr]
\cr}
\eqno(22)$$
and
$$\eqalign{{d\over {dK_m}}~ln &\left({{\xi_{1n}-T_m \xi_{1n}}\over
{\xi_{1n}-T_m\xi_{2n}}}
{{\xi_{2m}-T_m \xi_{2n}}\over {\xi_{2n}-T_m\xi_{1n}}}\right)
{}~=~{{\xi_{1n}-T_m\xi_{2n}}\over {\xi_{1n}-T_m \xi_{1n}}}
{{\xi_{2n}-T_m\xi_{1n}}\over {\xi_{2n}-T_m\xi_{2n}}}
\cr
&~~~~~~~~~~~~~~~~~~~~~~~~~~~~~~~~~~~~~~~{d\over
{dK_m}}\left[{{\xi_{1m}-T_m\xi_{1n}}\over {\xi_{1n}-T_m\xi_{2n}}}
{{\xi_{2n}-T_m\xi_{2n}}\over {\xi_{2n}-T_m\xi_{1n}}}\right]
\cr
{d\over{dK_m}}(\xi_{1n}-T_m\xi_{1n})~&=~{{(\xi_{1n}-\xi_{1m})
(\xi_{1n}-\xi_{2m})(\xi_{2m}-\xi_{1m})}\over
{[(\xi_{1n}-\xi_{2m})-K_m(\xi_{1n}-\xi_{1m})]^2}}
\cr
{d\over {dK_{m_j}}}T_{m_1}...T_{m_l}(\xi_{1n})&~=~
T_{m_1}^\prime (T_{m_2}...T_{m_l}(\xi_{1n}))T_{m_2}^\prime(T_{m_3}...T_{m_l}
(\xi_{1n}))~...
\cr
&~~~~~~~~~~~~~~~~~~~~T_{m_{j-1}}^\prime (T_{m_j}...T_{m_l}(\xi_{1n}))
\cr
\cdot
&{{(T_{m_{j+1}}...T_{m_l}(\xi_{1n})-\xi_{1m_j})
(T_{m_{j+1}}...T_{m_l}(\xi_{1n})-
\xi_{2m_j})(\xi_{1m_j}-\xi_{2m_j})}
\over
{[(T_{m_{j+1}}...T_{m_l}(\xi_{1n})-\xi_{2m_j})
-K_{m_j}(T_{m_{j+1}}...T_{m_l}(\xi_{1n})
-\xi_{1m_j})]^2}}
\cr
when~T_{m_{j+1}},~...,~T_{m_l}~&\ne~T_{m_j}
\cr}
\eqno(23)$$
the dependence of the following metric components on the multipliers can be
verified
$$g_{K_n,{\bar K}_n} \sim {1\over {\vert K_n\vert^2}}~~~~
g_{K_n, {\bar \xi}_{1n}} \sim {1\over {K_n}}~~~~
g_{K_n, {\bar \xi}_{2n}} \sim {1\over {K_n}}~~~~
g_{\xi_{1n}, {\bar K}_n} \sim {1\over {{\bar K}_n}}~~~~
g_{\xi_{2n}, {\bar K}_n} \sim {1\over {{\bar K}_n}}
\eqno(24)$$
Thus, to leading order, the multiplier part of the integration measure at genus
$g$ constructed from such a metric will be
$${{d^2K_1}\over {\vert K_1\vert^2}}~ {{d^2K_2}\over {\vert K_2\vert^2}}
... ~{{d^2K_g}\over {\vert K_g\vert^2}}~ {1\over {[det(Im~\tau)]^{g+1}}}
\eqno(25)$$
This resembles part of the superstring measure [12] and finiteness of
superstring amplitudes [13][14] is consistent with finiteness of the symplectic
volume of ${\cal H}_g/Sp(2g;{\Bbb Z})$ [5][15].

It is of particular interest to compare this measure at genus twelve with
the
\hfil\break
Polyakov measure resulting from the Mumford isomorphism of bundles over
moduli space
\hfil\break
$K\simeq E^{13}$ [16].  Using the reggeon calculus formalism [17], a moduli
space
measure has been constructed
$$\eqalign{\prod_{n=1}^g {{d^2K_n}\over {\vert K_n\vert^4}} \vert 1-K_n\vert^4&
{}~{1\over {Vol(SL(2,{\Bbb C}))}} \prod_{m=1}^g {{d^2\xi_{1m} d^2\xi_{2m}}\over
{\vert \xi_{1m}-\xi_{2m}\vert^4}}~[det(Im~\tau)]^{-13}
\cr
&\prod_\alpha~^\prime\prod_{p=1}^\infty \vert 1-K_\alpha^p\vert^{-48}
\prod_\alpha~^\prime \vert 1-K_\alpha\vert^{-4}
\cr}
\eqno(26)$$
which is consistent with the singularity and harmonic properties
characteristic of the Polyakov measure [1].  Equivalence of (26)
with the Polyakov measure at genus two and three, expressed in terms of theta
functions in equation (3), and consequently modular invariance of the
two- and three-loop reggeon measure, has been established [4].  However, as a
direct proof of equivalence at higher genus has not been feasible, because of
the absence of a similar representation in terms of period matrices, modular
invariance of the measure (26) at higher genus remains to be shown.

Nevertheless, at genus
twelve, the Polyakov measure differs from the measure induced by the symplectic
metric (18),
$d\mu_{12}^{symp}$, containing the terms in equation (25),
by a factor $\vert F_1\vert^{-2}$, which is singular at the
boundary of moduli space.  Since the reggeon measure (26) has the same
singularity properties as the Polyakov measure in the limit of  degenerate
Riemann surfaces, it will differ from $d\mu_{12}^{symp}$ by  the same
factor $\vert F_1(K_n, \xi_{1n}, \xi_{2n})\vert^{-2}$ up to the square of
the absolute value of an analytic function $\vert F_2(K_n, \xi_{1n}, \xi_{2n})
\vert^2$, defined on a subset of moduli space ${\cal M}_{12}^0$ [5].

Now consider a modular transformation of the measure (26).  The singularity
properties of the measure and therefore $\vert F_1\vert^{-2}$ will be
unchanged.  The transformed reggeon measure is given by the product of
$d\mu_{12}^{symp}$ and $\vert F_1\vert^{-2} \vert {\tilde F}_2\vert^2$, as
$d\mu_{12}^{symp}$ is modular invariant.  Since the ratio $\left\vert{{{\tilde
F}_2
(K_n, \xi_{1n}, \xi_{2n})}\over {F_2(K_n, \xi_{1n}, \xi_{2n})}}\right\vert^2$
can be regarded as the square  of the absolute value of an analytic function
on all of moduli space, spanned by the Schottky group coordinates, it may be
set equal to one after a suitable normalization of the measure.  This
argument suggests, therefore, that the expected modular invariance of the
moduli space measure derived from the reggeon formalism is valid at genus
twelve.

In conclusion, at genus two and three, there are two isomorphisms
and $K\simeq E^{13}$ and $K\simeq E^{g+1}$ giving rise to two different
modular invariant measures (3) and (17)  respectively, involving the
period matrix coordinatization of moduli space.  Modular invariance of the
two- and three-loop measures (3) follows because the factor arising in the
transformation $\prod_{i\le j}~d\tau_{ij}$ is only an extra phase.  Even in the
absence of an isomorphism $K\simeq E^{g+1}$ at higher genus, a
modular-invariant measure
 (17), obtained from the symplectic metric in the Siegel upper half space
${\cal H}_g$, can be re-expressed in terms of Schottky group parameters
and compared directly with the reggeon measure.

\vskip .5in

\centerline{\bf Acknowledgements.}
I would like to thank Prof. S. W. Hawking and Dr. G. W. Gibbons for their
encouragement while this research has been undertaken.
I also wish to acknowledge useful discussions with K. S. Narain, K. Roland
and A. Sen about
two-loop string physics.
\vfill
\eject

\centerline{\bf REFERENCES}

\item{1.}  A. Belavin and V. Knizhnik, ${\underline{Phys.~Lett.}}$
 ${\underline{\underline {B168}}}$
(1986) 201 - 206
\item{2.}  G. Moore, ${\underline{Phys.~Lett.}}$
 ${\underline{\underline {B176}}}$ (1986) 369 - 379
\item{3.}  A. Belavin, V. Knizhnik, A. Morozov and A. Peremelov,
${\underline{Phys.~Lett.}}$
${\underline{\underline {B177}}}$ (1986) 324 - 328
\item{4.} K. O. Roland, ${\underline{Nucl.~Phys.}}$
 ${\underline{\underline{B313}}}$ (1989) 432 - 446
\item{5.}  S. Nag, ${\underline{Proc.~Indian~Acad.~Sci.~(Math.~Sci.)}}$
${\underline{\underline {99}}}$(2)
(1989) 103 - 111
\item{6.}  S. Davis, ${\underline{Class.~Quantum~Grav.}}$
 ${\underline{\underline 7}}$ (1990) 1887 - 1893
\item{7.}  B. A. Dubrovin, ${\underline{Russian~Math.~Surv.}}$
 ${\underline{\underline {36}}}$ (1981) 11 - 80
\item{8.}  M. Mulase, ${\underline{J.~Diff.~Geom.}}$
 ${\underline{\underline {19}}}$ (1984) 403 - 430
\item{9.}  T. Shiota, ${\underline{Inv.~Math.}}$
 ${\underline{\underline {83}}}$ (1986) 333 - 382
\item{10.}  M. Matone, Imperial College preprint, IC-Math/9-92
\item{11.}  S. Davis, University of Cambridge preprint, DAMTP-R/94/15
\item{12.}  P. DiVecchia, K. Hornfeck, M. Frau, A. Lerda and S. Sciuto,
${\underline{Phys.~Lett.}}$
 ${\underline{\underline {B211}}}$ (1988) 301 - 307
\item{13.}  N. Berkovits, ${\underline{Nucl.~Phys.}}$
 ${\underline{\underline{B408}}}$ (1993) 43 - 61
\item{14.}  S. Davis, University of Cambridge preprint, DAMTP-R/94/27
\item{15.}  C. L. Siegel, ${\underline{Symplectic~Geometry}}$ (New York:
Academic Press, 1964)
\item{16.}  D. Mumford, ${\underline{L.~Ens.~Math.}}$
 ${\underline{\underline {24}}}$ (1977) 39 - 110
\hfil\break
P. Nelson, ${\underline{Phys.~Rep.}}$
${\underline{\underline {149C}}}$ (1987) 337 - 375
\item{17.}  J. L. Petersen and J. R. Sidenius, ${\underline{Nucl.~Phys.}}$
${\underline{\underline
{B301}}}$ (1988) 247 - 266

\end